\documentclass[prb,aps,,superscriptaddress,longbibliography,twocolumn]{revtex4-1}
\usepackage{tikz}  
\usetikzlibrary{shapes,arrows,positioning,automata,backgrounds,calc,er,patterns}

\usepackage[T1]{fontenc}
\usepackage{graphicx}
\usepackage{dcolumn}
\usepackage{bm}
\usepackage{amsmath}
\usepackage{amssymb}
\usepackage{amsthm}
\usepackage{amsfonts}
\usepackage{enumerate}
\usepackage{array}
\newcolumntype{P}[1]{>{\centering\arraybackslash}p{#1}}
\usepackage{float}

\usepackage{lipsum}

\usepackage{braket}

\usepackage[colorlinks=true,urlcolor=blue,citecolor=blue,allcolors=blue]{hyperref}
\usepackage{cleveref}

\begin{document}

\title{Complexity of fermionic states}
\author{Tuomas I. Vanhala}
\affiliation{Computational Physics Laboratory, Physics Unit, Faculty of Engineering and
Natural Sciences, Tampere University, P.O. Box 692, FI-33014 Tampere, Finland}
\affiliation{Helsinki Institute of Physics P.O. Box 64, FI-00014, Finland}

\author{Teemu Ojanen}\email{Email: teemu.ojanen@tuni.fi}
\affiliation{Computational Physics Laboratory, Physics Unit, Faculty of Engineering and
Natural Sciences, Tampere University, P.O. Box 692, FI-33014 Tampere, Finland}
\affiliation{Helsinki Institute of Physics P.O. Box 64, FI-00014, Finland}

\begin{abstract}
How much information a fermionic state contains? To address this fundamental question, we define the complexity of a particle-conserving many-fermion state as the entropy of its Fock space probability distribution, minimized over all Fock representations. The complexity characterizes the minimum computational and physical resources required to represent the state and store the information obtained from it by measurements. Alternatively, the complexity can be regarded a Fock space entanglement measure describing the intrinsic many-particle entanglement in the state. We establish universal lower bound for the complexity in terms of the single-particle correlation matrix eigenvalues and formulate a finite-size complexity scaling hypothesis. Remarkably, numerical studies on interacting lattice models suggest a general model-independent complexity hierarchy: ground states are exponentially less complex than average excited states which, in turn, are exponentially less complex than generic states in the Fock space. Our work has fundamental implications on how much information is encoded in fermionic states.

\end{abstract}
\maketitle

\section{Introduction}

The complexity of an object or a process quantifies how something can be generated from simple building blocks in an optimal way. For example, the computational complexity of a mathematical operation is defined in terms of the number of elementary operations required in its execution, or the complexity of a unitary operation in a quantum computer is defined as a the minimum number of elementary quantum gate operations required in its generation.\cite{nielsen2006,nielsen_chuang_2010} Various notions of complexity in quantum systems and their relation to quantum information processing has been actively studied recently.\cite{susskind1,chapman, PhysRevD.106.046007} In this work, we introduce the complexity of $N$-particle fermionic states. The complexity quantifies how resource intensive it is to express a given state as a linear combination of Slater states (fermionic product states), which are the building blocks of the fermionic Fock space. If the complexity of a state is $\mathcal{C}$, to express this state in any Fock basis, one needs to specify at least $\mathcal{C}$ nonzero coefficients. We show rigorously that the Fock representation of a state cannot be compressed to less than $n_{\mathrm{qubits}}=\log_2 \mathcal{C}$ qubits, showing how the complexity determines the minimal physical and computational resources required to represent the state. Sophisticated numerical methods\cite{schollwock1,orus1} have been developed to mitigate the exponential complexity of correlated systems, however, only a genuine quantum simulation\cite{Feynman} can be expected to incorporate it in general. Our results provide quantitative estimate for the complexity of distinct classes of states, outlining required resources for the quantum simulation targeting fermionic states.\cite{gonzales-cuadra,BRAVYI2002210,abrams,ortiz}

\begin{figure}[b]
\centering
\includegraphics[width=0.85\columnwidth]{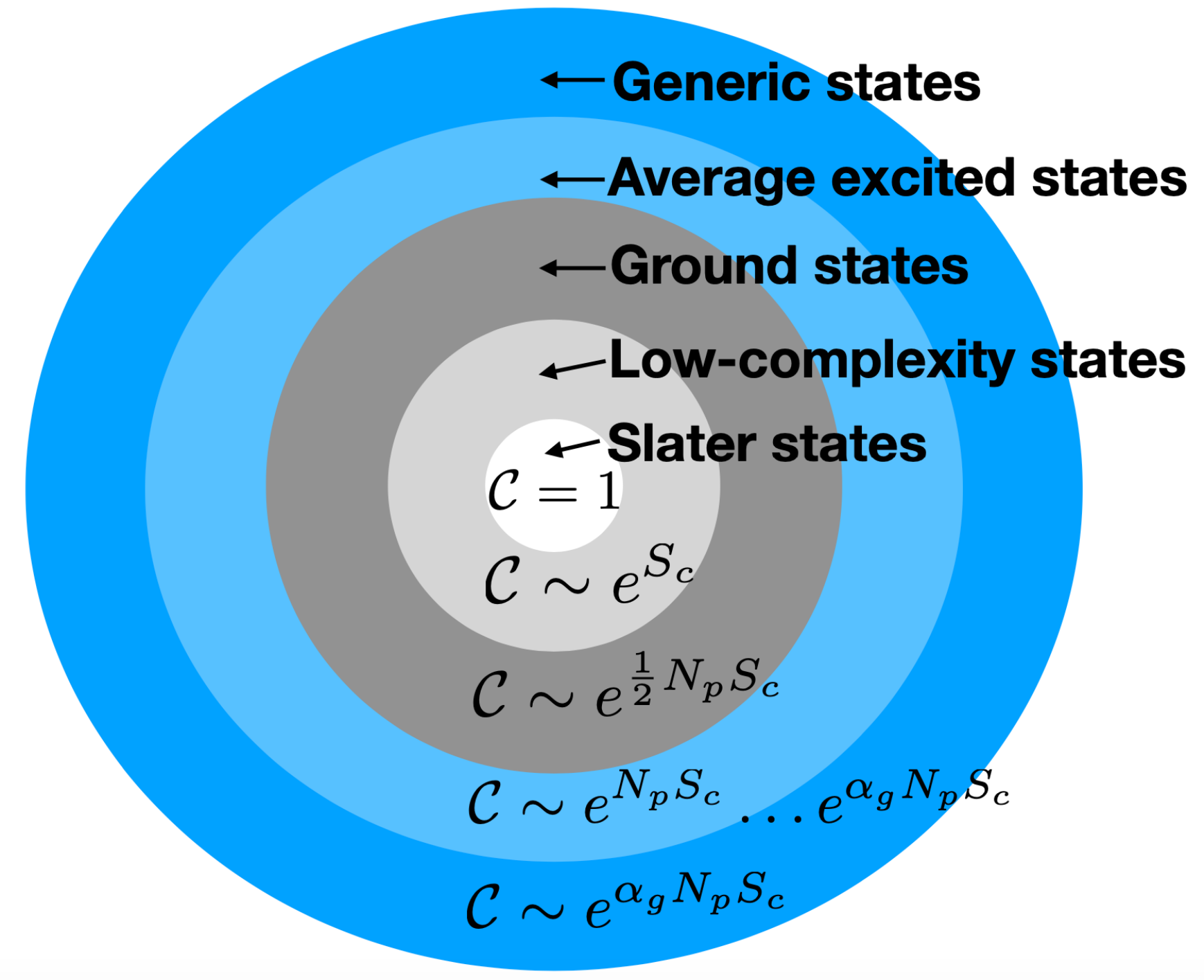}
\caption{Complexity hierarchy of fermionic states as a function of particles $N_p$ and the correlation entropy $S_c$ at fixed filling fraction. Ground states and excited states refer to eigenstates of interacting lattice Hamiltonians at strong coupling exceeding the bandwidth and $1\leq\alpha_g\leq 2$, depending on the filling.  }
\label{fig1}
\end{figure}

Besides its computational and information-theoretic implications, the complexity constitutes an entanglement measure in the Fock space. In contrast to widely studied partition entanglement measures,\cite{amico} the complexity describes intrinsic partition-independent properties of $N$-particle states. It sharply distinguishes between the states of interacting and noninteracting Hamiltonians: all non-degenerate eigenstates of noninteracting systems can be represented as a single Slater state, thus having a trivial complexity.

The central finding in our work is that the complexity for distinct classes of states can be faithfully estimated from the correlation entropy $S_c$, defined essentially as the entanglement entropy between a single particle and the rest of the system. The quantity $S_c$, exhibiting intensive size scaling, is calculated from the eigenvalues of the single-particle correlation matrix (i.e. the natural occupations), and thus easily available in many numerical and theoretical methods. By relating the complexity $\mathcal{C}$ to $S_c$, we thus quantify and give a precise meaning to the idea, put forward already in the seminal works of L\"{o}wdin,\cite{lowdin1} that the natural occupations can be used to estimate how far the state is from a single Slater configuration.\cite{Collins+1993+68+74,PhysRevA.54.259,PhysRevA.103.042816,PhysRevC.107.044318} Furthermore, this relates $\mathcal{C}$ as a newly defined entanglement measure to the widely-studied entanglement entropies of the $N$-particle reduced density matrices.\cite{PhysRevA.92.042326,BENATTI20201,PhysRevA.102.042410,PhysRevA.103.052424} Specifically, i) we establish a universal lower bound for the complexity  $S_P\geq S_c$, where $S_P$ is the logarithmic complexity $\mathcal{C}=e^{S_P}$ ii) we introduce a model-independent finite size complexity scaling hypothesis $S_P\sim \alpha N_p S_c$ for homogeneous $N_p$-particle states with constant filling fraction iii) numerical studies of interacting lattice models suggest that the coefficient $\alpha$ characterizes universal features of distinct classes of states, implying the exponential complexity hierarchy summarized in Fig.~1. In strongly coupled lattice models, the ground states are exponentially less complex than average excited states, which in turn are exponentially less complex than the generic states in the Fock space. Due to the model-independent nature of the scaling hypothesis, we postulate that the same complexity scaling is applicable for a broad class of local Hamiltonians. Our work has fundamental implications on how much information is contained in fermionic states.

\section{Fermionic complexity}

We begin by defining the complexity for an arbitrary fermionic state $|\Phi\rangle$ in the Fock space of $N_p$ identical particles and $N_o$ available single-particle orbitals. This state can be expanded as  
\begin{align}\nonumber
|\Phi\rangle=\sum_{k=1}^{k_{max}} a_{\{n_{B_i} \}_k}|\{n_{B_i} \}_k\rangle  \end{align}
where $B_i$ denotes orbital $i$ in the single-particle basis $\mathcal{B}$, and $\{n_{B_i} \}_k$ labels the distinct sets of single particle occupation numbers $n_{B_i}=0,1$. The $N_p$-particle Slater basis states are defined as $|\{n_{B_i} \}_k\rangle=\hat{c}^\dagger_{B_{jN_p}}\ldots\hat{c}^\dagger_{B_{j2}}\hat{c}^\dagger_{B_{j1}}\ket{0}$, where the product of fermion creation operators contains the populated orbitals in the set $\{n_{B_i}\}_k$. Each Slater state is multiplied with a nonzero complex probability amplitude $a_{\{n_{B_i} \}_k}\neq 0$. Depending on $|\Phi\rangle$ and the employed single-particle orbitals $\mathcal{B}$, the number of terms $k_{max}$ varies between 1 and the Fock space dimension $Q=\binom{N_o}{N_p}$. We now consider the 2nd Renyi entropy of the probability distribution of the Slater states 
\begin{align}\nonumber
 S_{P_\mathcal{B}}=-\ln{\sum_{k} P_{k}^2 }, 
\end{align}
where $P_{k}=|a_{\{n_{B_i} \}_k}|^2$ denotes the probability weight of $|\{n_{B_i} \}_k\rangle$ in $|\Phi\rangle$. To eliminate the dependence on $\mathcal{B}$, we define \emph{the logarithmic complexity} as
\begin{align}\label{minimize}
 S_P=\min_{\mathcal{B}}S_{P_\mathcal{B}}, 
\end{align}
where the minimization is carried over all possible single-particle bases $\mathcal{B}$. Finally, we define the complexity of the state $|\Phi\rangle$ as
\begin{align}\nonumber
\mathcal{C}=e^{S_P}.
\end{align}
In practical calculations, carrying out the minimization in Eq.~\eqref{minimize} is highly nontrivial task. Remarkably, as seen below, for the eigenstates of the studied lattice Hamiltonians, the optimal basis is excellently approximated by the correlation matrix eigenbasis and the position basis at weak and strong coupling.  

The complexity, as defined above, has two illuminating interpretations: \textbf{i)} The complexity of a state determines its maximum compression in the Fock space, characterizing the number of terms in the most compact representation. By employing fundamental results in classical and quantum information theory\cite{shannon,schumacher,preskill2015lecture}, we show in App.~\ref{app:compression} that the maximum compression of the quantum information in a fermionic state is determined by its complexity. Specifically, we prove that the number of qubits required to encode the Fock space information of a state is, at least, $n_{\mathrm{qubit}}=S_P\log_2e$. This characterizes the minimum physical resources required to represent and store general fermionic many-body states. We emphasize that the result, which has close parallels with Shannon's and Schumacher's encoding theorems in classical and quantum information theory, is universal and applies to generic quantum simulation and quantum information platforms. \textbf{ii)} The complexity of a state describes its intrinsic $N$-particle entanglement. Without entanglement, the state could be represented as a single Slater state. If the state has complexity $\mathcal{C}$, the amount of entanglement corresponds to that in an equal superposition of $\mathcal{C}$ Slater states. Different aspects of Fock space (or mode) entanglement, has been studied extensively over the years. \cite{PhysRevA.92.042326,BENATTI20201,PhysRevA.102.042410,PhysRevA.103.052424,PhysRevA.67.024301,RevModPhys.81.865} However, none of the previous mode entanglement measures capture the same information as the complexity studied here. In fact, the complexity is unique in establishing a concrete connection between the information content of the state and the $N$-particle entanglement. In contrast to the entanglement entropy and other partition-based measures, the complexity does not depend on arbitrary case-specific partition. Moreover, the complexity sharply distinguishes interacting and non-interacting systems since all non-degenerate eigenstates of quadratic Hamiltonians can be represented as a single Slater state with $S_P=0$, irrespective whether they obey the area-law\cite{eisert1}, the volume-law\cite{bianchi} or the critical entanglement entropy scaling.

The complexity $S_P$ can be contrasted with other quantities derived from coefficients in the Fock basis. The \emph{Slater rank}\cite{PhysRevA.64.022303,ECKERT200288} of a state is the minimal number of Slater determinants required to exactly expand the state, and has been studied in low-dimensional systems.
This is related to the generalized Pauli constraints, \cite{altunbulak2008pauli,PhysRevLett.110.040404} as exactly satisfying a constraint can lead to lower-dimensional representation of the state even if the average occupations do not take values of $0$ or $1$ (c.f. the discussion on $S_c$ below), although this may again be mostly relevant in low-dimensional systems.\cite{reuvers2021generalized} Measures similar to the Slater rank have also been considered in bosonic systems. \cite{PhysRevLett.116.080402,chin2018generalized}
Another extreme is to consider only the weight of the largest Slater determinant, \cite{PhysRevA.89.012504,PhysRevA.94.032513} which was considered as an entanglement measure for the Laughlin wave function.\cite{2017NJPh...19h3019Z}

The complexity $\mathcal{C}$, the Slater rank and the largest weight are related to Renyi-$n$-entropies of the Slater weights with $n=2$, $n=0$ and $n=\infty$. Indeed, the generalization of the complexity for other Renyi entropies and the Shannon entropy is immediate, and the relation to the corresponding single-particle entanglement entropies discussed below in the $n=2$ case will be developed in a forthcoming work for general $n$. The Renyi-2 entropy has the advantage of an elementary proof of the lower bound property discussed below and admitting certain analysis tools\cite{PhysRevB.103.214206} used in App.~\ref{sec:hamming_dist}, while the Shannon entropy has a more direct relation to asymptotic information content of the state, as discussed in App.~\ref{app:compression}.

\subsection{Complexity lower bound from natural orbitals} \label{sec:complexity_lower_bound}

A central role in the complexity is played by the single-particle correlation matrix, also known as the one-body reduced density matrix,
\begin{align} \nonumber
C_{ij}=\langle\Phi|\hat{c}^\dagger_{j}\hat{c}_{i}|\Phi\rangle,
\end{align} 
where $\hat{c}^\dagger_{i}, \hat{c}_{j}$ denote the fermionic creation and annihilation operators and indices $i,j \in 1 \dots N_o$ label all possible single-particle orbitals in a fixed basis. If we have a system with $N_o$ available orbitals, the correlation matrix has dimension $N_o\times N_o$. Due to Fermi statistics, the correlation matrix eigenvalues satisfy $0\leq\lambda_i\leq1$ and $\sum_i\lambda_i=N_p$. Thus, they can be interpreted as single-orbital occupation probabilities in the eigenbasis of $C_{ij}$.
The eigestates of $C_{ij}$ are commonly referred as the natural orbitals\cite{lowdin1}, which have found modern applications in analyzing strongly correlated many-body systems.\cite{boguslawski1,aikebaier1} We can define one-particle correlation entropy in the state $|\Phi\rangle$ as
\begin{align} \nonumber
S_{c}^p=- \ln{\frac{\sum_i\lambda_i^2}{N_p}},
\end{align}
which is a measure of how the occupation probabilities of the natural orbitals collectively differ from 1 or 0. As discussed in App~\ref{app:corr}, up to a trivial constant, $S_c^p$ is equal to the Renyi entanglement entropy between a single particle and the rest of the system. Generic properties of entanglement entropies of fermionic N-particle reduced density matrices have been discussed extensively,\cite{PhysRevA.92.042326,BENATTI20201,PhysRevA.102.042410,PhysRevA.103.052424} and quantities similar to $S_c^p$ have been employed to characterize phase transitions in many-particle systems\cite{PhysRevB.105.115145,PhysRevLett.115.046603} and as measures of correlation\cite{lowdin1,PhysRevA.54.259,PhysRevA.103.042816,Collins+1993+68+74} and complexity. \cite{PhysRevC.107.044318} It is thus useful to establish a connection between $S_P$ and $S_c$, clarifying the role of $S_P$ as a novel entanglement measure.

By interchanging the role of particles and holes, we define single-hole occupation probabilities $\tilde{\lambda}_i=1-\lambda_i$, which satisfy $0\leq \tilde{\lambda}_i\leq1$ and $\sum_i\tilde{\lambda}_i=N_o-N_p$. We then define a single-hole correlation entropy as 
\begin{align} \nonumber
S_{c}^h=- \ln{\frac{\sum_i\tilde{\lambda}_i^2}{N_o-N_p}},
\end{align} 
and \emph{the correlation entropy} as the larger of the two 
\begin{align}
S_{c}=\mathrm{max} \left\{ S_{c}^p, S_{c}^h\right\}.
\end{align} 
In App~\ref{app:proof} we prove that, for arbitrary fermionic state $|\Phi\rangle$, the correlation entropy provides a lower bound for the logarithmic complexity
\begin{align}\label{eq:bound}
 S_P\geq S_{c}.
\end{align}
This complexity bound is nontrivial: there exist states with nonzero logarithmic complexity for which the lower bound is saturated. A simple example is obtained by considering states where the number of particles $N_p$ and available orbitals $N_o$ satisfy $nN_p\leq N_o$ with $n\geq 2$. Given a single-orbital basis, one can find at least $n$ disjoint occupation number sets $\{n_{B_i} \}_k$ where each occupied orbital with $n_{B_i}=1$ belongs precisely to one set. Forming a superposition of such occupation sets $|\psi\rangle=\sum_{k=1}^{n} \sqrt{P_k}|\{n_{B_i} \}_k\rangle$, with $\sum_k P_k=1$, yields an example of a complexity bound saturating state. For these states the correlation matrix $C$ is diagonal, and the natural occupations $\lambda_i=C_{ii}=P_k$ for each of the $N_p$ occupied orbitals in the set $\{n_{B_i} \}_k$. Thus, $S_c=S_p=-\ln \sum_i^n P_i^2$, and the lower bound in \eqref{eq:bound} is saturated. The state $|\psi\rangle$ can also be regarded as an $n$-orbital generalization of the Greenberger-Horne-Zeilinger state $\frac{1}{\sqrt{2}}\left(|\uparrow \uparrow \uparrow\ldots\rangle+|\downarrow\downarrow\downarrow \ldots\rangle\right)$.\cite{PhysRevA.88.012335} These type of states, whose complexity do not scale with the total number of particles at fixed filling fraction $\nu=\frac{N_p}{N_o}$, define the low-complexity category in Fig.~1. This category include, for example, eigenstates of impurity systems with a non-extensive number of scattering centers, such as the Kondo model. Despite a macroscopic reorganization of the Fermi sea, the eigenstates have only a few correlation matrix eigenvalues that differ from 0 or 1.\cite{debertolis}

\subsection{Complexity scaling}

\label{sec:complexity_scaling}

The existence of the lower-bound \eqref{eq:bound} saturating states suggests that the bound cannot be significantly improved without making additional assumptions of the states of interest. Eigenstates of local interacting Hamiltonians and other large-scale homogeneous states defined on a $d$-dimensional spatial lattice constitute a class of central importance. They define a family of states which can be studied as a function of the system size for a fixed filling fraction $\nu$. How is the complexity of such states scaling as the system size grows? For a generic filling fraction $\nu\neq 0,1$, the dimension of the Fock space of such states grows exponentially in $N_p$. Thus, in the leading order, we expect that the logarithmic complexity scales as $S_P\sim N_p$. However, the maximum value of the correlation entropy does not scale with the system size $S_c\leq \max\{-\ln \nu, -\ln(1-\nu)\}$. This shows that $S_c$ alone does not provide an accurate approximation for the complexity of these states. However, the role of $S_c$ in Eq.~\eqref{eq:bound} suggests that it encodes some universal features of the complexity. Combining this idea with the exponential scaling in the system size, we postulate that the complexity of uniform states follows, in the leading order, the scaling form
\begin{align}\label{eq:scaling}
 S_P\sim \alpha N_i S_{c},
\end{align}
where $\alpha>0$ captures universal features of distinct classes of states. Here $N_i$ is the number of particles $N_i=N_p$ when $\nu\leq \frac{1}{2}$, and the number of holes $N_i=N_o-N_p$ when $\nu>\frac{1}{2}$.  We illustrate this hypothesis for three paradigmatic examples: the Hubbard model of spinful fermions 
\begin{align}\nonumber
\hat{H}=t\sum_{\langle i, j\rangle ,\sigma}\left(\hat{c}^\dagger_{i\sigma}\hat{c}_{j\sigma}+\mathrm{h.c.} \right)+U\sum_{\langle i, j\rangle,\sigma}\hat{n}_{i\sigma}\hat{n}_{i\bar{\sigma}},
\end{align}
the $t-V$ model of spinless fermions 
\begin{align} \nonumber
\hat{H}=\sum_{\langle i, j\rangle}\left(t\hat{c}^\dagger_{i}\hat{c}_{j}+ \mathrm{h.c.}+V\hat{n}_{i}\hat{n}_{j}\right),
\end{align}
and Haar-distributed states, which we call ``generic states'' as they represent uniformly distributed unit vectors in the Fock space (see Sec.~D in Methods). We observe that, indeed, the value of $\alpha$ distinguishes different broad classes of states:
 \begin{enumerate}
     \item {The generic states have $\alpha=\alpha_g$, where $1\leq\alpha_g\leq 2$, depending on the filling fraction. The maximum $\alpha_g=2$ is obtained at $\nu=\frac{1}{2}$, while $\alpha_g\to1$ when $\nu\to 0$ or $\nu\to 1$.}
    \item {For non-degenerate ground states, $\alpha=\frac{1}{2}$ provides an excellent lower bound, which can become tight in various limits.}
    \item { Average excited states have $1\lesssim\alpha<\alpha_g$ when the interaction exceeds the bandwith.} 
 \end{enumerate}
The difference in $\alpha$, despite its innocent appearance, translates into an exponential difference in the complexity. 
\begin{figure*}
\includegraphics[width=0.95\linewidth]{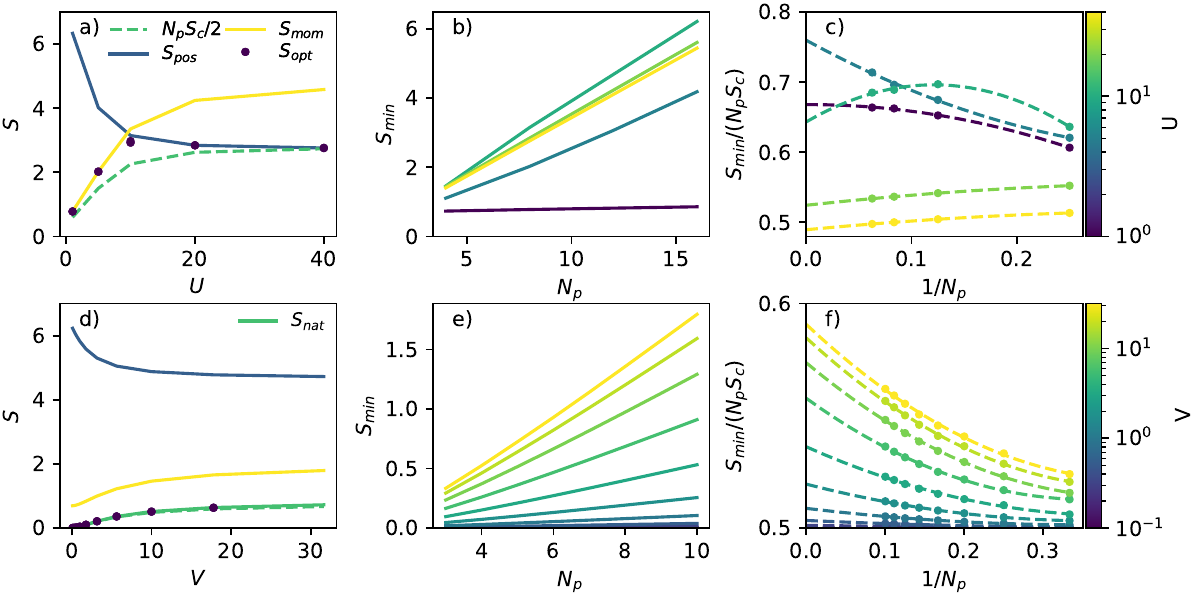}
\caption{Complexity of ground states. a): Ground state complexity of a half-filled Hubbard chain as a function of interaction. The system length is $L=8$ sites ($N_o=16$ orbitals). The numerically optimized complexity (black dots) is well-approximated from below by the ground state scaling form $S_P=\frac{1}{2}N_pS_c$ (green dashed line). The solid lines correspond to complexities calculated in the momentum and position orbitals. Complexity at weak coupling does not approach zero due to ground state degeneracy in the non-interacting model. b): Complexity scaling of the half filled Hubbard chain as a function of the particle number $N_p$. The complexity $S_P$ is approximated by $S_{min}=\min(S_{pos},S_{mom})$, as we cannot perform the full optimization for large systems. c): Tentative extrapolation of the coefficient $\alpha=S_P/(N_pS_c)$ to infinite system size. The dashed lines are a quadratic fit. At strong coupling the value approaches $\alpha \approx 1/2$. d)-f): The same quantities for the $t-V$ model but at $\nu=1/3$ filling. Here we also explicitly include the complexity $S_{nat}$ in natural orbitals, which always gives the approximated minimal complexity $S_{min}$. Again, the ground state complexity follows closely the scaling form $S_P=\frac{1}{2}N_pS_c$.}
\label{fig2}
\end{figure*}
The complexity of generic states provides a baseline reference to compare other types of states. The analytical expression for $\alpha_g$ is derived in App~App~\ref{app:generic}. The generic states saturate the maximum value of the correlation entropy $S_c$ and the maximal leading order complexity allowed by the dimensionality of the Fock space. As seen below, the eigenstates of local Hamiltonians allow exponential compression compared to the generic states.

In Fig.~\ref{fig2} we illustrate the ground state complexity for the Hubbard model for $\nu=\frac{1}{2}$ and the $t-V$ for $\nu=\frac{1}{3}$. The minimizing basis, found by the conjugate gradient optimization (see  App~\ref{app:detail} for details), is well approximated by the momentum states at weak coupling. In this case, the momentum basis is a natural orbital basis, however, the natural orbitals are not unique due to degeneracies in the natural occupations. For the $t-V$ model, the natural orbitals are essentially the optimal basis also at strong coupling, while for the Hubbard chain, the optimal basis at strong coupling coincides with the position orbitals. The ground state complexity for both models is seen to satisfy $S_P\gtrsim \frac{1}{2}N_pS_c$, where the lower bound appears tight for small $V$ and large $U$. In the Hubbard chain, the correlation entropy saturates the maximum value $S_c=\ln 2$ at strong coupling. Thus the logarithmic complexity of a generic state at half filling, $S_P=2N_p\ln 2$, is four times larger than that of the ground state of the Hubbard chain $S_P \approx \frac{1}{2}N_p\ln 2$ at strong coupling. Furthermore, the size scaling suggests that $\alpha$ converges reasonably close to $\alpha=\frac{1}{2}$ for all coupling strengths. This is observed for both models at fillings for which the ground state is non-degenerate. For the $t-V$ model at half filling, the ground state corresponds to two near-degenerate charge density wave configurations. In this case, we observe that the complexity of each charge-density wave state is well-captured by $S_P\sim\frac{1}{2}N_pS_c$. We discuss the mechanism leading to the specific value $\alpha=1/2$ in App.~\ref{sec:hamming_dist}, and provide further data for other filling fractions in App.~\ref{sec:additional_ground_state_data}.

\begin{figure*}
\centering
\includegraphics[width=0.9\linewidth]{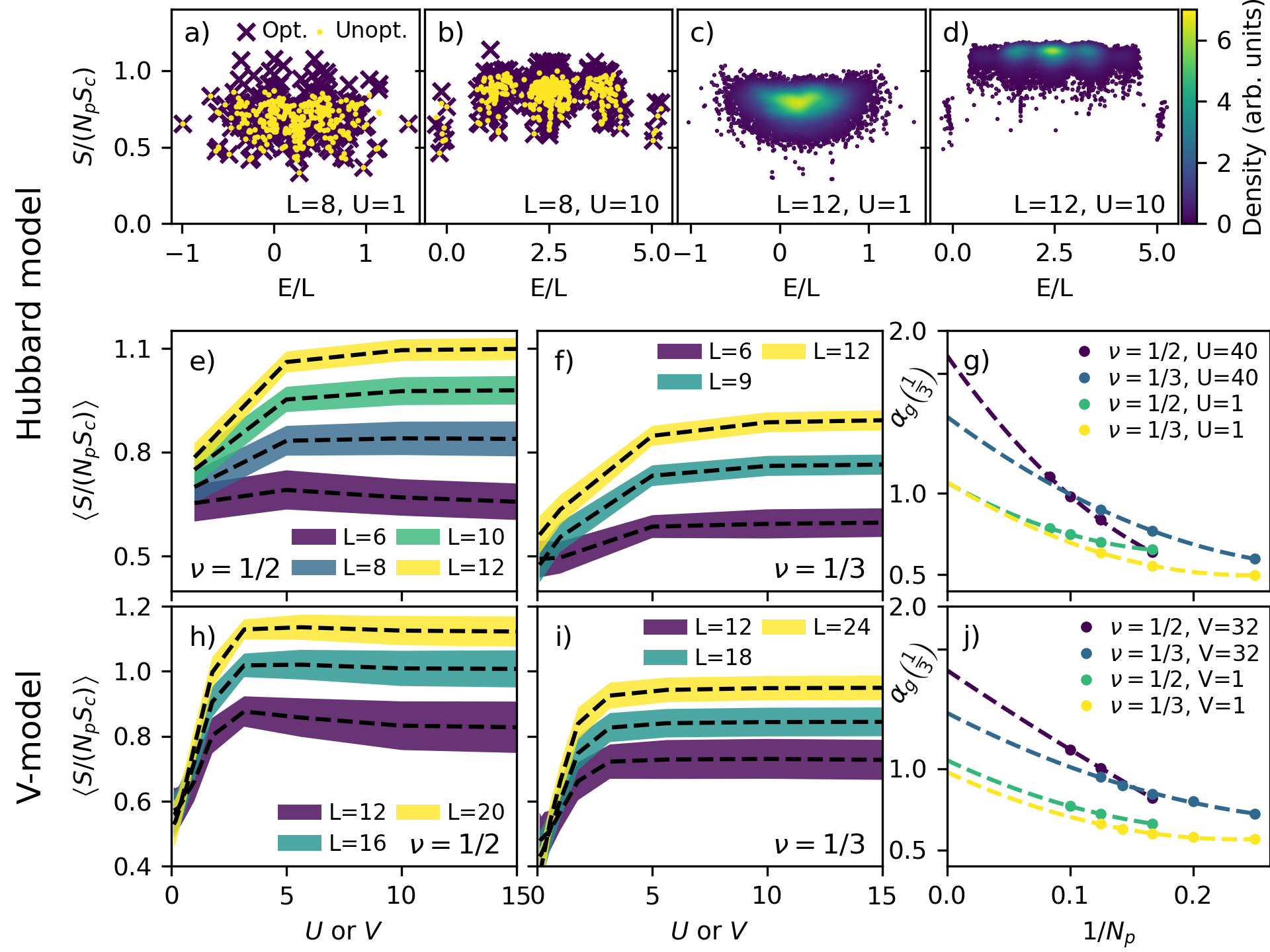}
\caption{Complexity of excited states in the Hubbard and the $t-V$ model in the ground state parity and momentum sector. a)-b): Comparison of numerically optimized results $S_{opt}$ with $S_{min}=\min(S_{nat},S_{mom},S_{pos})$, where $S_{nat}$, $S_{mom}$ and $S_{pos}$ are the natural orbital, momentum and position basis complexities in a Hubbard chain of length $L=8$. c)-d): Distribution of the excited state complexity ratio $S/(N_p S_c)$ for a larger system of $L=12$. e)-f): Mean complexity ratio in the ground state sector of the Hubbard model for filling factors $\nu=1/2$ and $\nu=1/3$. The colored areas around the mean (dashed line) have a width of one standard deviation. g): Scaling of the mean complexity with the number of particles $N_p$. For comparison, generic states have $\alpha_g=2$ at $\nu=1/2$ and $\alpha_g$ at $\nu=1/3$ is marked on the axis. h)-j): the same as e)-f) but for the $t,V$ model instead of the Hubbard model.}
\label{fig3}
\end{figure*}

In Fig.~\ref{fig3} we analyze the complexity of excited states, for the same systems as in Fig.~\ref{fig2}, by performing a full diagonalization in the parity and center-of-mass momentum sector which contains the ground state. For the excited states, finding numerically the minimizing basis becomes more challenging. As seen in Figs.\ref{fig3} a)-b), the numerical optimization does not find the true minimum for some high-complexity states. However, in the vast majority of cases, the optimization converges very close to the minimum value over natural, momentum and position orbitals. This indicates that, like for the ground states, one obtains an accurate approximation for the complexity by analyzing only these orbitals, especially when considering averages over many states. For both models, the average complexity of the excited states grows as a function of interaction and saturates to a constant at $U/t,V/t\approx 4$. At strong coupling, the average complexity is substantially higher than for the ground states. While the full diagonalization is restricted to modest system sizes, a fact one should be conscious of in extrapolating the results, Fig.~3 g) and f) imply that the ratio $S_P/(N_pS_c)$ for the average excited states converge to a constant $\alpha<\alpha_g$. The specific value of $\alpha$ depends on the coupling strength and filling, but the average excited states remain, even around the midspectrum, significantly less complex than generic states. This behaviour is markedly different from the entanglement entropy, which exhibits identical leading order volume-law scaling for the midspectrum states of nonintegrable Hamiltonians and generic states. \cite{vidmar1,bianchi,kliczkowski} We also note that the quantity $S_P/\log(Q)$, where $Q$ is the number of basis states, can be seen as a basis-independent multifractal coefficient, which is connected to  quantum ergodicity and thermalization. We present an outlook on this connection in App.~\ref{sec:multifractal_outlook}.

\section{Discussion}
In the above, we have seen how the complexity hierarchy summarized in Fig.~1 emerges. The single-Slater states, such as the eigenstates of quadratic Hamiltonians, have trivial complexity and are regarded as the fundamental building blocks of more complex states. For the low-complexity states, for which the complexity is not scaling with the system size when filling fraction is  fixed, the complexity can be estimated from the universal lower bound $S_c$. As seen above, the ground state complexity is typically well captured by the scaling Ansatz \eqref{eq:scaling} with prefactor $\alpha=1/2$. When the interaction exceeds the bandwidth, the complexity of average excited states follow \eqref{eq:scaling} with $1\lesssim \alpha<\alpha_g$, where the upper bound determines the complexity of generic states. The model-independent nature of the scaling hypothesis and the qualitative agreement of different models suggest that the above results are not sensitive to the specific form of the Hamiltonian, as long as some broad features, such as locality and large scale homogeneity, are satisfied.

\section{Conclusion and outlook}

We introduced the complexity of a fermionic state to quantify the amount of information in it. The complexity provides a bound to the quantum state compression by choosing an optimal Fock basis, determining the minimum computational and physical resources to represent states. We showed that, for distinct classes of states, the complexity can be estimated from the eigenvalues of the single-particle correlation matrix. Considering the rapidly increasing interest in fermionic quantum simulation and quantum information processing, our results open several topical avenues of research.  Does the observed complexity scaling laws for ground states and excited states represent a fundamental limit in encoding information to the eigenstates of local Hamiltonians? Do the complexity scaling laws, as their model-independent form suggests, also hold for higher dimensional systems? How can the scaling laws for eigenstates be derived from general arguments? To what extent the discovered complexity structure applies to bosonic states? How does the notion of Fock complexity, as studied here, reflect the circuit complexity of concrete fermionic quantum simulation schemes?\cite{gonzales-cuadra,PhysRevLett.127.020501} Here we especially want to mention the concept of ``magic'',\cite{PRXQuantum.3.020333} a property of the quantum states critical to speedup over classical computation, and the fact that all non-gaussian fermionic states can be considered to possess ``magic''.\cite{PhysRevLett.123.080503} Answers to these questions would provide fundamental new insight in many-body systems and their quantum information applications.

\acknowledgements  T.O. acknowledges the Academy of Finland (project 331094) and Jane and Aatos Erkko Foundation for support. Computing resources were provided by CSC -- the Finnish IT Center for Science.

\appendix

\section{ Complexity as a fundamental bound to quantum state compression } \label{app:compression}

Here we illustrate two key aspects of the complexity: its role as the minimum resources required to store the information of a many-body state and characteristic number of terms in the optimal Fock representation.    

\subsection{Complexity and the optimal many-body compression rate}
In information theory, the notion of entropy was introduced to quantify the compression of strings of data which follows a known distribution.\cite{shannon} Analogously, the logarithmic complexity, which is an entropy quantity, characterizes the compression of the quantum information in a many-body state. Here we provide a derivation of this fundamental property. Following similar steps as in the entanglement distillation,\cite{nielsen_chuang_2010} we consider an optimal encoding of $n$ copies of a fermionic many-body state $|\Phi\rangle=\sum_{k} a_{\{n_{B_i} \}_k}|\{n_{B_i} \}_k\rangle $, where $\{n_{B_i} \}_k$ denotes an occupation number set in the single-orbital basis $\mathcal{B}$, and $k\in\{1,2,\ldots Q\}$ where $Q$ is the Fock space dimension. Thus, the object of interest is a composite state
\begin{align}
  &|\Psi\rangle=\prod_{j=1}^{n} \otimes|\Phi\rangle=  |\Phi\rangle\ldots\otimes|\Phi\rangle\otimes|\Phi\rangle=\nonumber 
  \\
  &\sum_{k_1\ldots k_n} \sqrt{P_{k_1}^{\mathcal{B}}P_{k_2}^{\mathcal{B}}\ldots P_{k_n}^{\mathcal{B}}}|\{n_{B_i} \}_1\rangle\otimes|\{n_{B_i} \}_2\rangle\ldots |\{n_{B_i} \}_n\rangle,  \nonumber
\end{align}
 which is an element of $Q^n$-dimensional composite Fock space. In the above, the probabilities are defined as $P_{k_i}^{\mathcal{B}}=|a_{\{n_{B_i} \}_k}|^2$ and the complex phases of amplitudes are absorbed in the Fock basis states. In general, composite states $|\Psi\rangle$ live in a lower-dimensional subspace $\mathcal{H}$ of $Q^n$. Schumacher's encoding theorem implies that, in the limit of large $n$, state $|\Psi\rangle$ can be projected into a typical subspace of dimension $2^{nH(P_k^{\mathcal{B}})}$ with arbitrary high accuracy, where $ H(P_ k^{\mathcal{B}})= -\sum_k P_ k^{\mathcal{B}}\log_2 P_ k^{\mathcal{B}} $ is the Shannon entropy.\cite{shannon,preskill2015lecture} The projection operator into the $\delta$-typical subspace is of the form  
\begin{align}
  P(\delta,n)=\sum_{\delta-\mathrm{typical}} |k_1\rangle\langle k_1|\otimes|k_2\rangle\langle k_2|\ldots |k_n\rangle\langle k_n|,  \nonumber
\end{align}
where the $\delta$-typical states are defined by $|P_{k_1}P_{k_2}\ldots P_{k_n}-2^{-nH(P_ k^{\mathcal{B}})}|\leq \epsilon$, and number of such states is at most $2^{n(H(P_ k^{\mathcal{B}})+\delta)}$. For arbitrary $\epsilon,\delta>0$, it is always possible to achieve  
\begin{align}
  ||P(\delta,n)|\Psi\rangle||=1-\epsilon,  \nonumber
\end{align} 
by allowing for sufficiently large $n$.\cite{nielsen_chuang_2010} This implies that
\begin{align}
  &|\Psi\rangle=\sum_{k_1\ldots k_n} \sqrt{P_{k_1}P_{k_2}\ldots P_{k_n}}|k_1\rangle\otimes|k_2\rangle\ldots |k_n\rangle=\nonumber \\
  &\sum_{\delta-\mathrm{typical}} \sqrt{P_{k_1}P_{k_2}\ldots P_{k_n}}|k_1\rangle\otimes|k_2\rangle\ldots |k_n\rangle+\mathcal{O}(\epsilon) = \nonumber\\
  &2^{-nH(P_ k^{\mathcal{B}})/2}\sum_{\delta-\mathrm{typical}}|k_1\rangle\otimes|k_2\rangle\ldots |k_n\rangle+\mathcal{O}(\epsilon).  \nonumber
\end{align}
Thus, for sufficiently large $n$, the state can be compressed into $2^{n H(P_ k^{\mathcal{B}})}$ dimensional subspace $\mathcal{H}$. The maximum compression is obtained in the Fock basis $\mathcal{B}$ which minimizes $H(P_ k^{\mathcal{B}})$. Since the Shannon entropy is bounded from below by the second Renyi entropy $H(P_ k^{\mathcal{B}})\geq -\log_2 \sum_k (P_ k^{\mathcal{B}})^2$, the minimum $H(P_ k^{\mathcal{B}})$ is bounded by the logarithmic complexity and
\begin{align} \label{eq:compress}
\ln (\dim \mathcal{H})\geq  nS_P, 
\end{align}
or, equivalently, $\dim \mathcal{H}\geq \mathcal{C}^n$.
This result has fundamental importance in the quantum information applications of many-body physics. The dimension of $\mathcal{H}$ should be regarded as the physical resource, and the asymptotic cost, required in storing $n$ copies of $|\Phi\rangle$. Thus, to represent and store the quantum information in $n$ copies of $|\Phi\rangle$ requires at least $N_{\mathrm{qubit}}=\log_2 \mathcal{C}^n=\log_2 e^{nS_P}$ qubits, or $n_{\mathrm{qubit}}=N_{\mathrm{qubit}}/ n=\log_2 e^{S_P}$ qubits per copy. To properly appreciate the fundamental nature of the results, we recall Shannon's classical result which states that a string of $n\gg 1$ letters, each appearing with probability $P_k$, can be optimally compressed to $n_{\mathrm{bit}}=\log_2 2^{nH(P_k)}$ bits. This shows that the logarithmic complexity has a similar role in encoding quantum information of many-body states what the Shannon entropy has in encoding classical information.\footnote{Had we defined the complexity in terms of the minimized Shannon entropy of the Fock distribution, the analogy would be even more exact. However, working with the Renyi entropy is technically convenient in the present work, and the complexity still provides the lower bound for the number of qubits required to represent the state.} In summary, the complexity characterizes the minimum resources to represent many-body states in the Fock space and underlines the physical requirements of all quantum simulation and quantum information applications.

\subsection{Complexity and quantum information from measurements}

In addition to characterizing the optimal information compression, the complexity also characterizes the information that can be obtained from a many-body state by measurements. This section can be regarded as a complementary way to understand the formal result \eqref{eq:compress} in more physical terms. Let's consider that we prepare multiple copies of state $|\Phi\rangle$ and perform repeated  $N_p$-particle measurements in some Fock basis $|\{n_{B_i} \}_k\rangle$, where $\{n_{B_i} \}_k$ denotes an occupation number set in the single-orbital basis $\mathcal{B}$, and $k\in\{1,2,\ldots Q\}$ where $Q$ is the Fock space dimension. The resulting quantum states, obtained as outcomes of the $n$ measurements, constitutes the total information obtained from the measurements. This information can be stored as a composite state of the form  
\begin{align}\label{eq:measure}
|\{n_{B_i} \}_{k_1}\rangle\otimes |\{n_{B_i} \}_{k_2}\rangle\otimes \ldots \otimes|\{n_{B_i}, \}_{k_n}\rangle,
\end{align}
which is an element of $Q^n$-dimensional Hilbert space. Again we will see that the complexity of  $|\psi\rangle$ provides a fundamental lower bound of how much of the $Q^n$-dimensional space such states cover. In the language of quantum information theory, these composite states, obtained with probability $P_{k_1}^{\mathcal{B}}P_{k_2}^{\mathcal{B}}P_{k_3}^{\mathcal{B}}\ldots P_{k_n}^{\mathcal{B}}$, can be regarded as quantum messages constructed from individual letters, where each letter is a quantum state drawn from the ensemble  $\left\{ |\{n_{B_i} \}_{k}\rangle,P_{k}^{\mathcal{B}}\right\}$. Now one can ask how much these quantum messages can be compressed, or what is the minimum dimension of space $\mathcal{H}$ in which the messages can be accurately stored when $n$ is large. The dimension of $\mathcal{H}$ determines the physical resources needed to store information extracted from $|\psi\rangle$. This formulation turns the problem into an application of Schumacher's encoding theorem\cite{schumacher,preskill2015lecture} in the special case where the letters form an orthogonal set. In this case, the quantum state of the messages is uniquely indexed by strings $k_1k_2\ldots k_n$. When $n$ is large, Shannon's noiseless coding theorem implies that these strings can be faithfully compressed to $2^{n H(P_ k^{\mathcal{B}})}$ long strings, where $ H(P_ k^{\mathcal{B}}) $ is the Shannon entropy.\cite{shannon,preskill2015lecture} Thus, in this limit, almost all messages fit into a space of dimension $\log_2 \left(\dim \mathcal{H}\right)=nH(P_ k^{\mathcal{B}})$. The maximum compression is obtained in the Fock basis that minimizes $H(P_ k^{\mathcal{B}})$ bounded by the logarithmic complexity $\ln (\dim \mathcal{H})\geq  nS_P$ in agreement with Eq.~\eqref{eq:compress}.  Thus, the complexity provides a lower bound to the dimension of $\mathcal{H}$, where the states obtained by $n$ measurements can be stored. Whenever the logarithmic complexity is smaller than its maximum value $\ln Q$, the composite states obtained from $|\psi\rangle$ by $n$ measurements do not fill the whole $Q^n$ dimensional space exhaustively but only a subspace of it.

\subsection{Complexity as the characteristic number of terms in the minimal Fock representation}

In addition to its rigorous role as an optimal quantum information compression rate discussed above, the complexity of a state is also connected to the characteristic number of terms which are required to span it in the optimal basis. Let $\{P_n\}$ be the probabilities in the optimal Fock basis which determines the complexity and let's assume that the distribution is arranged in non-increasing order $P_{n_1}\geq P_{n_2}$ when $n_1<n_2$. How many terms are needed in the optimal basis to effectively span the state? Specifically, how large should $\tilde{n}$ be to satisfy $\sum_{n=1}^{\tilde{n}} P_n\sim 1$? This question is important for the states with large complexity $\tilde{n},\mathcal{C}\gg 1$ and the answer depends on the distribution: i) for sufficiently uniform distributions with a well-defined typical probability scale, the required number of terms is $\tilde {n}\sim \mathcal {C}$  ii) for heavy-tailed distributions, the required number of terms can scale nonlinearly in the complexity $\tilde{n}\sim \mathcal{C}^\beta$ with $\beta>1$.  

Let's first study  i) and consider a case where the probabilities have a characteristic order of magnitude $P_n\sim P_0$ when $n\leq n'$, and are strongly suppressed for $n>n'$. This implies that $P_0\sim 1/n'$ and $\mathcal {C}=1/\sum_n P^2\sim n'$. Thus, the complexity roughly coincides with the effective cutoff index $n'$ and we can conclude that $\sum_{n=1}^{\tilde{n}} P_n\sim 1$ is achieved when $\tilde{n}\sim \mathcal{C}$. When the distribution is strictly box distribution with constant probabilities $P_0=1/n'$, the full probability is exactly recovered after $\mathcal{C}$ terms $\sum_{n=1}^{\mathcal{C}} P_n= 1$. In general, to recover the full probability for distributions with a finite tail above $n=n'$, one might need to include a few multiples of $\mathcal{C}$ terms. The linear scaling between $\tilde{n}$ and $\mathcal{C}$ reflects the typical expectation that entropy-like quantities scale as the logarithm of the total number of contributing states.  

\begin{figure}
\centering
\includegraphics[width=0.9\linewidth]{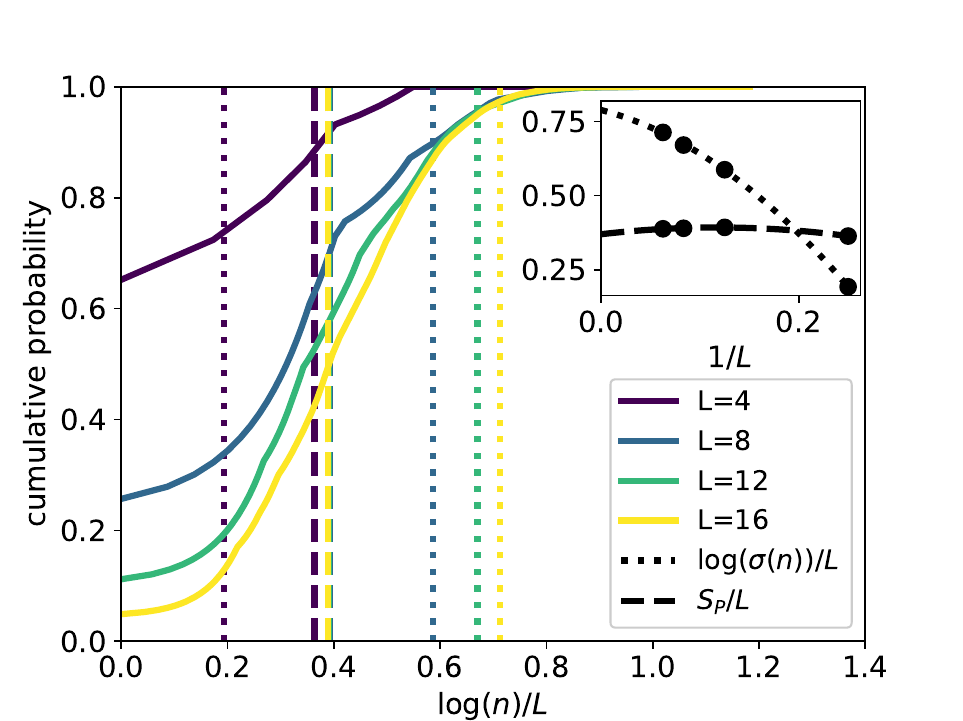}
\caption{Cumulative probability distribution for the ground state of the Hubbard model at $U=10$, as a function of the scaled logarithm of the state index $\log(n)/L$. The state probabilities $P_n$ have been ordered from the largest to smallest. The vertical lines mark the scaled entropy $S_P/L$ and $\log(\sigma(n))/L$, where $\sigma(n)$ is the standard deviation of $n$. The inset shows a tentative extrapolation of $S_P/L$ and $\log(\sigma(n))/L$ to infinite system size.}
\label{fig_n_distribution}
\end{figure}

In case ii), the distribution has a long tail, the probabilities do not have a well-defined scale, and the previous reasoning breaks down. For this type of distributions, a nonlinear scaling $\tilde{n}\sim \mathcal{C}^\beta$ with a model-specific $\beta>1$ becomes possible. Such behaviour can be observed, for example, for power-law distributions and the distributions of eigenstates of lattice Hamiltonians, as illustrated in Fig.~\ref{fig_n_distribution}.  For the ground state of a strongly coupled Hubbard model, we find that  $\sum_{n=1}^{\mathcal{C}} P_n\sim 0.5$ and that the standard deviation and the complexity of the optimal distribution satisfy $\sigma=C^\beta$, where $\beta\lesssim 2$. Because most of the probability is located withing a few standard deviations, the full probability is covered by $\sum_{n=1}^{m\mathcal{C}^\beta} P_n\sim 1$, where $m$ is a small integer.

To summarize, the complexity of a state provides a lower bound estimate for the characteristic number of terms in the optimal Fock representation.

\section{Correlation entropy as an entanglement entropy} \label{app:corr}

The single-particle correlation matrix in state $\left|\Phi\right\rangle$ is conventionally defined as $C_{ii'}=\langle\Phi|\hat{c}^\dagger_{i'}\hat{c}_{i}|\Phi\rangle$, which can also be written in first quantized notation as
\begin{equation*}
C_{ii'}=N_p \sum_{j,k,l,...} \Phi(i,j,k,l,...) \Phi(i',j,k,l,...)^*,
\end{equation*}
where $\Phi(i,j,k,l,...)$ is the antisymmetric wave function of the particles at coordinates $i,j,k,l,...$~.\cite{lowdin1} Thus the actual normalized reduced density matrix of a single particle, defined as the partial trace over the coordinates of the other particles, is $\rho_1=C/N_p$,\cite{PhysRevB.105.115145} and the order $n$ Renyi entanglement entropy is defined as
\begin{equation*}
S_n=\frac{1}{1-n}\ln(\mathrm{Tr}(\rho_1^n)),
\end{equation*}
and has the von Neumann limit $S_1=-\mathrm{Tr}(\rho_1\log(\rho_1))$.

If $\left| \Phi \right\rangle$ is a single Slater determinant, $C$ has $N_p$ times degenerate eigenvalue $1$, the rest being zero. Therefore the entanglement entropies become
\begin{equation*}
S^{\text{Slater}}_n=-\frac{1}{1-n}\log(N_p^{n-1}).~\text{(Slater state)}
\end{equation*}
However, in the spirit of the complexity $S_P$ which is trivial for Slater states, we subtract the free fermion contribution and define the single-particle \emph{correlation entropies} as
\begin{equation}
S_{c,n}=S_n-S^{\text{Slater}}_n=\frac{1}{1-n}\ln\left(\mathrm{Tr}\left( \frac{C^n}{N_p} \right)\right).
\end{equation}
$S_{c,2}$ is the particle correlation entropy discussed in the main text. Thus, the correlation entropy is actually a one-particle entanglement entropy from which the free fermion contribution has been subtracted.



\section{Proof of the complexity lower bound} \label{app:proof}

Here we will give a proof of the complexity lower bound \eqref{eq:bound} in three steps.

\textbf{Proposition 1}: Let's consider a fermionic $N_p$ particle state $|\Phi\rangle$. Furthermore, let's assume that $\lambda_i$ is the set of correlation matrix eigenvalues (occupation probabilities of the natural orbitals) and $\bar{n}_{B_i}$  are the occupation probabilities of single-particle orbitals in an arbitrary basis $\mathcal{B}$. They always satisfy $\sum_i\lambda_i^2\geq \sum_i{\bar{n}_{B_i}}^2$. \textbf{Proof}: Let $C$ be the correlation matrix. Then  $\sum_i\lambda_i^2=\mathrm{Tr}\,C^2=\sum_{\alpha,\beta}C_{\alpha\beta}C_{\beta\alpha}=\sum_{\alpha,\beta}|C_{\alpha\beta}|^2\geq \sum_{\alpha}|C_{\alpha\alpha}|^2\equiv\sum_{\alpha}{\bar{n}_{B_\alpha}}^2$. Here we used the fact that the in the double sum all entries are positive and that occupation probabilities in a general basis are defined as diagonal entries of the correlation matrix. 

\textbf{Proposition 2}: The average occupation numbers $\bar{n}_{B_i}$  and the state probabilities $P_{\{n_{B_i} \}_k}$ always satisfy $\sum_i\frac{\bar{n}_{B_i}^2}{N_p}\geq\sum_{k}P_{\{n_{B_i} \}_k}^2$. The first sum is over all the single-particle orbitals whereas the second sum is over all $k_{max}$ occupation number sets in Eq.~(1). \textbf{Proof}: The average occupation numbers can be written in terms of the state probabilities as $\bar{n}_{B_i}=\sum_{k}P_{\{n_{B_i} \}_k}n_{B_i}^k$, where $n_{B_i}^k=0,1$ is the value of the occupation number of orbital $B_i$ in the set $\{n_{B_i}\}_k$. From this we get $\sum_i\bar{n}_{B_i}^2=\sum_i\sum_{k,l}P_{\{n_{B_i} \}_k}n_{B_i}^k P_{\{n_{B_i} \}_l}n_{B_i}^l\geq \sum_i\sum_{k}P_{\{n_{B_i} \}_k}^2n_{{B_i}_k}^2= \sum_{k}P_{\{n_{B_i} \}_k}^2\sum_i n_{{B_i}_k}^2=\sum_{k}P_{\{n_{B_i} \}_k}^2\sum_i n_{B_i}^k=\sum_{k}P_{\{n_{B_i} \}_k}^2N_p$. The inequality follows from dropping non-negative terms from the double sum.  Comparing the starting and final form, we have proved Proposition 2.

\textbf{Universal lower bound for $S_P$}: using Property 1. and the monotonicity of logarithm, we deduce that $S_{c}^p=-\log{\frac{\sum_i\lambda_i^2}{N_p}}\leq -\log{\frac{\sum_i\bar{n}_{B_i}^2}{N_p}} $. Now, using Property 2 it follows that  $S_{c}^p\leq -\log{\frac{\sum_i\bar{n}_{B_i}^2}{N_p}}\leq -\log{\sum_{k}P_{\{n_{B_i} \}_k}^2}=S_{P_{\mathcal{B}}} $.
Since this holds for arbitrary basis $\mathcal{B}$, we can minimize the right-hand side over all bases and it still holds. Thus, we have proved that $S_{c}^p\leq S_P.$ The corresponding inequality for the hole correlation entropy  $S_{c}^h\leq S_P$ can be straightforwardly established by exchanging the roles of particles and holes and tracing the same steps. Thus, we  arrive at  $S_{c}\leq S_P$ where $S_{c}$ is the larger one of $S_{c}^p,S_{c}^h$.

\section{ Complexity of generic states } \label{app:generic}

\begin{figure}
\centering
\includegraphics[width=0.9\linewidth]{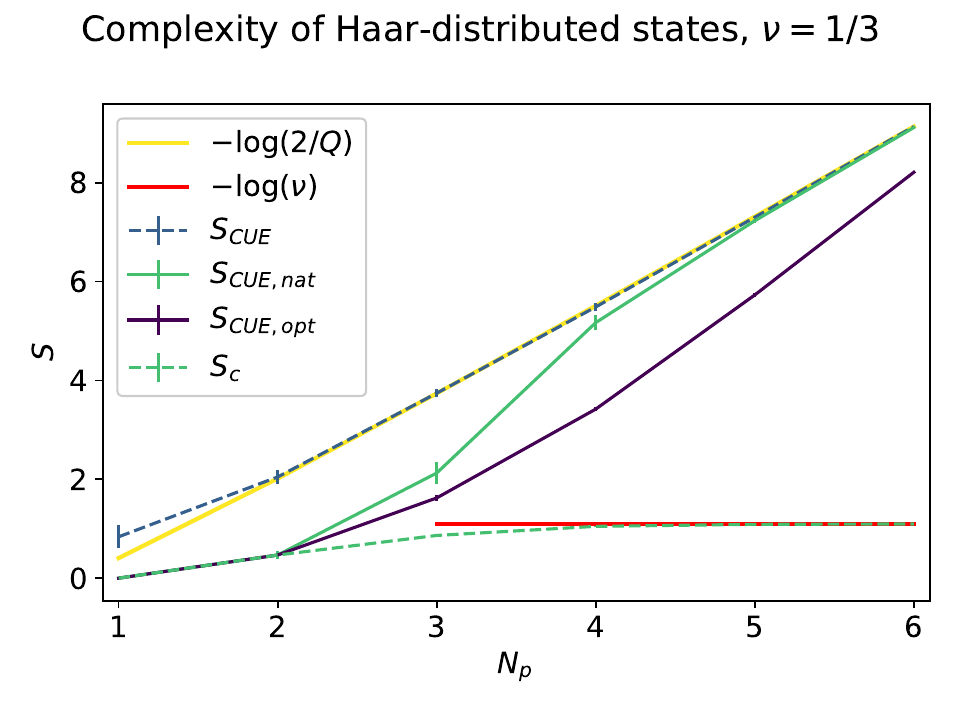}
\caption{Complexity of generic, Haar-distributed states in different single-particle bases computed as a mean of five samples with the vertical bars giving the sample standard deviation, except at $N_p=6$ where we have only computed one realization. A Q-dimensional generic state has logarithmic complexity $S_{CUE}=-\log(2/Q)$ in the limit of large $Q$. The complexity can be reduced by applying single-particle rotations as demonstrated here by transforming to the natural orbital basis or a numerically optimized basis. However, to conserve information, we expect that the typical reduction in the characteristic number of terms $C=\exp(-S)$ in the optimal Fock representation of a state with no special structure scales with the number of parameters in the single-particle basis, i.e. polynomially in $N_p$, while $C$ grows exponentially for a fixed filling fraction. $S_{CUE,nat}$ and $S_{CUE,opt}$ are thus expected to approach $S_{CUE}$ for large $N_p$, while the correlation entropy approaches its maximal value $-\log(\nu)$.}
\label{fig4}
\end{figure}

Here we derive the complexity of generic states in a Fock space with $M$ available orbitals and $N_p$ particles with dimension $Q=\binom{N_o}{N_p}=\frac{N_o!}{(N_o-N_p)!N_p!}$. Let's start with some normalized vector in the Fock space 
$|\psi_0\rangle=\sum_{k=1}^Qa_k|k \rangle$,
where $|k \rangle$ is some basis and $\sum_k |a_k|^2=1$ and consider all the states that can be obtained from $|\psi_0\rangle$ by unitary transformations:
\begin{align}\nonumber
|\psi\rangle=U|\psi_0\rangle= \sum_{k,j=1}^QU_{jk}a_k|j \rangle.
\end{align}
These states fill the Fock space uniformly and are referred as generic states. To calculate the complexity, we extract the probabilities $P_j=|U_{jk}a_k|^2=U_{jk}U_{jl}^*a_ka_l^*$ (repeated indices are summed) and their squares $P_j^2=U_{jk}U_{jl}^*U_{jm}U_{jn}^*a_ka_l^*a_ma_n^*$. To evaluate the average $\langle P_i^2 \rangle$ over the Haar measure, we can make use of the circular unitary ensemble result\cite{log_gases} 
\begin{align}\nonumber
\langle U_{jk}U_{jm}U_{jl}^*U_{jn}^*  \rangle=\frac{1}{Q^2}\left(\delta_{kl}\delta_{mn}+\delta_{kn}\delta_{ml} \right)
\end{align}
for $Q\gg 1$.
Employing the above formula, we obtain $\langle P_j^2  \rangle=\frac{2}{Q^2}$,
and 
\begin{align}\nonumber
\sum_{j=1}^{Q}\langle P_j^2\rangle=\frac{2}{Q}
\end{align}
For large $Q\gg 1$, the average logarithmic complexity becomes $S_P=\langle -\ln{\sum_iP_i^2}\rangle=-\ln{\sum_i\langle P_i^2\rangle}=-\ln (2/Q)$. The expectation value can be moved inside the logarithm, because the argument becomes non-fluctuating in the large $Q$ limit. Also, the minimization over possible single-particle orbitals would not affect the result in large systems, since the number of optimization parameters scale linearly in orbitals while the independent components of the states vectors grow exponentially. The this behaviour is illustrated in Fig.~5, showing how the optimized complexity in small systems is approaching the above analytical results. By employing Stirling's formula, the leading order complexity of generic states state becomes  
\begin{align}\label{eq:generic1}
S_{P}= -N_o\ln\left[\nu^\nu(1-\nu)^{1-\nu}\right],
\end{align}
where $\nu=N_p/N_o$.  Since the generic states are uniformly distributed in space and cannot be compressed, their leading order complexity is the maximum allowed by the dimensionality of the Fock space. As illustrated in Fig.~5,  the generic states also maximize the particle and hole correlation entropies $S_c^p=-\ln \nu$, $S_c^h=-\ln (1-\nu)$. Thus, the result \eqref{eq:generic1} can be expressed in the general form \eqref{eq:scaling} with $\alpha=\alpha_g$ where
\begin{align}
\alpha_g&=1+\frac{(1-\nu) \ln (1-\nu)}{\nu \ln (\nu)},\qquad 0\leq\nu\leq \frac{1}{2}\nonumber\\
\alpha_g&=1+\frac{\nu \ln (\nu)}{(1-\nu) \ln (1-\nu)},\qquad  \frac{1}{2}<\nu\leq 1 \nonumber
\end{align}

\section{Outlook on multifractal coefficients and ergodicity} \label{sec:multifractal_outlook}

\begin{figure}
  \begin{center}
    \includegraphics[width=\columnwidth]{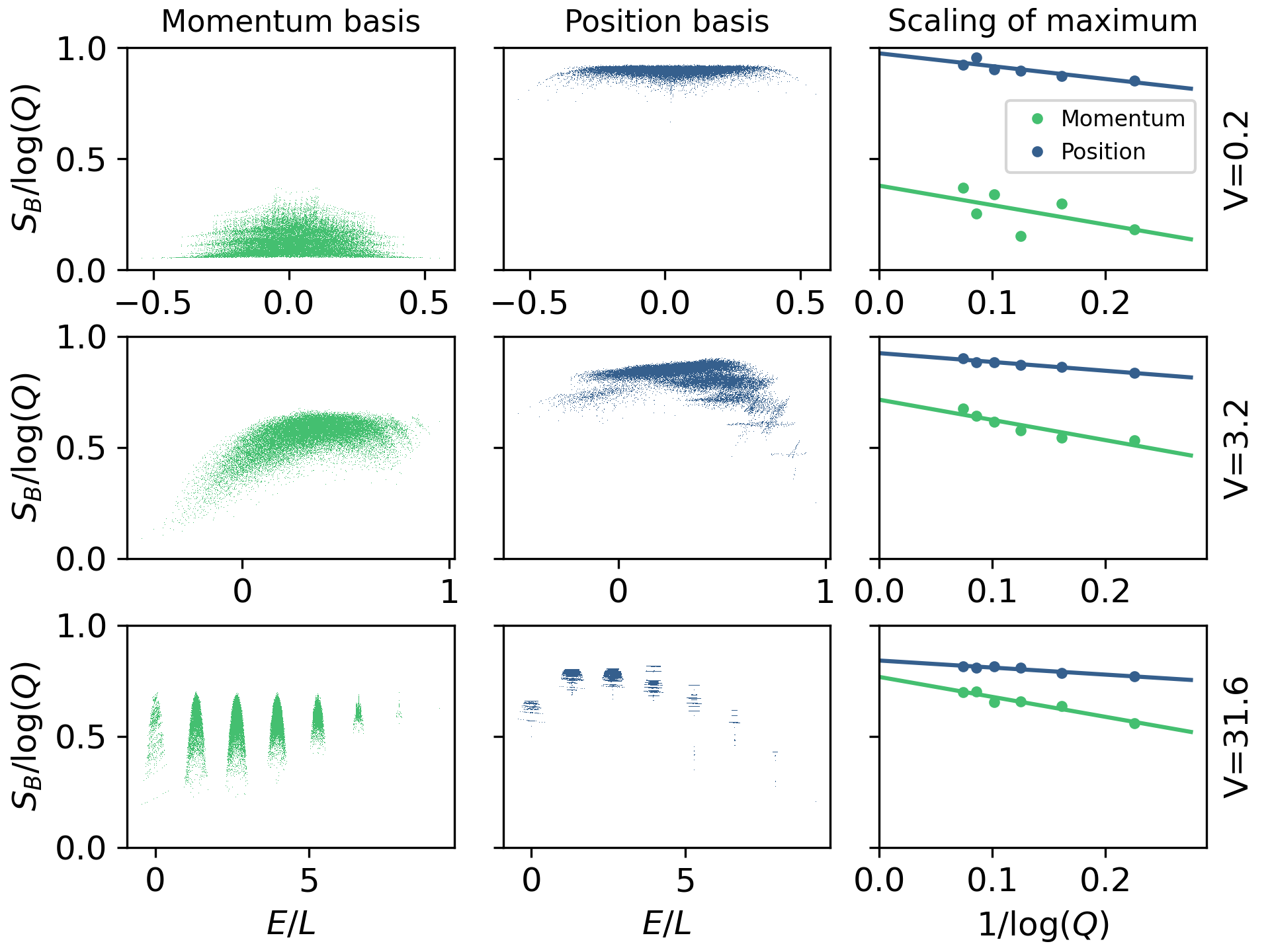}
    \caption{ Fractal dimension $D_{2,B}$ of the states in the ground state
      symmetry block of the $1/3$ filled $t - V$ model at different
      interaction strengths. The left and middle columns show the
      fractal dimension in the $L=24$ model in position and momentum
      bases. The right column shows the scaling of the maximal fractal
      dimension (i.e. the highest point in the left and middle panels)
      as a function of system size. $S_B$ is the Renyi-2 entropy of
      the Slater configuration distribution in the indicated basis and
      $Q$ is the number of states in the Fock
      space. \label{fig:fractal_coefficient_figure}}
  \end{center}
\end{figure}

The space-filling properties of quantum state vectors have been intensively studied in the context of the eigenstate thermalization hypothesis (ETH), quantum ergodicity and many-body localization \cite{PhysRevE.86.021104,PhysRevLett.123.180601,PhysRevLett.124.200602,PhysRevE.100.032117,PhysRevB.103.214206}. Given a fixed basis $B$, one can define multifractal coefficients which quantify the extent of the wave vector relative to the full basis size. For example, the Fock-basis multifractal coefficient \cite{PhysRevB.103.214206} $D_{2,B}=S_{P_B}/\log(Q)$, where $Q$ is the number of basis states, ranges from $D_{2,B}=0$ for a Slater state to $D_{2,B}=1$ for a uniformly distributed state. Using the complexity, one can then define the basis-independent quantity $D_2=S_P/\log(Q)$ which is $D_{2,B}$ minimized over the single-particle bases $B$. As an example, we plot $D_{2,B}$ for the $1/3$-filled $t - V$-model in Fig. \ref{fig:fractal_coefficient_figure}.

For chaotic spin models it has been demonstrated that midspectrum eigenstates are ``ergodic'': their fractal dimension approaches $1$ in the thermodynamic limit \cite{PhysRevLett.123.180601,PhysRevE.100.032117}. However, it is much less clear what fractal dimensions should be expected from a chaotic Hamiltonian when moving away from the center of the spectrum, as the states start to develop structure that may effectively limit the available basis states, potentially lowering the multifractal coefficient. If one considers assigning temperatures on the eigenstates based on subsystem density matrices, the midspectrum states are close to infinite temperature, while away from midspectrum the states have more structure and the temperature is finite.\cite{PhysRevE.107.024102} In analogy to this, the single-particle density matrix reveals structure in fermionic states that limits the complexity, and thus the minimal multifractal coefficients, as quantified by the scaling relation $S_P=\alpha N_p S_c$. Indeed, reaching $D_2=1$ is only expected if both $\alpha$ and $S_c$ approach the maximal, generic state values. As long as there is any one-particle structure and, thus, $S_c$ remains below the maximum, we expect to find $D_2<1$. For models with density-density interactions the eigenbasis of both the single-particle and the two-particle parts of the Hamiltonian is a Slater basis, and one may generically expect one-particle structure to be present even at midspectrum. Indeed, the basis-independent fractal dimension $D_2$ does not seem to reach $1$ for any interaction strength in Fig. \ref{fig:fractal_coefficient_figure}. More rigorous upper bounds for the Fock-space multifractal coefficients in terms of single-particle and higher correlations will be established in the future work.

\section{Additional data on ground state complexity} \label{sec:additional_ground_state_data}

\begin{figure}
  \begin{center}
    \includegraphics[width=\columnwidth]{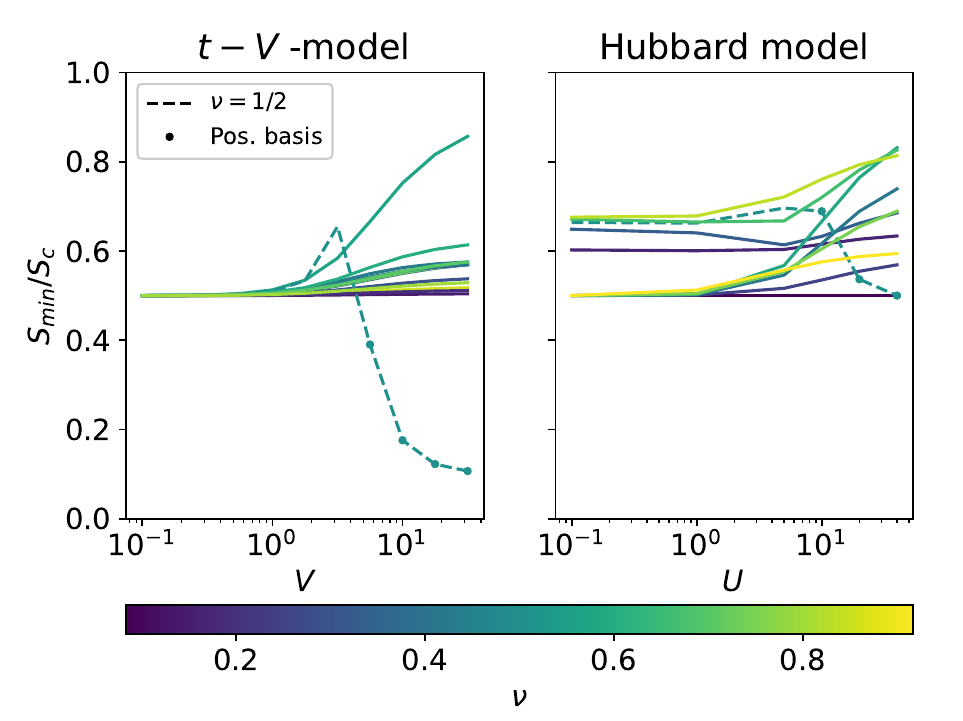}
    \caption{ The ratio $S_{min}/S_c$ for a range of filling factors $\nu$ in a Hubbard model of size $L=12$ and $t-V$-models of size $L=20,21$. $S_{min}$ is again the minimal $S_{P_B}$ chosen from the natural orbital, momentum and position bases. In some cases the ground state is degenerate, but the degeneracy can be resolved by the total momentum and reflection parity quantum numbers. We find that the complexity for the degenerate ground states is the same in all symmetry sectors. The basis giving the lowest complexity is typically the natural orbital basis in the $t-V$-model or the momentum basis in the Hubbard model, where the natural occupations may become degenerate. The exception is at half-filling (dashed lines) with strong interactions, where the position basis leads to lower complexity (marked with dots), and in the case of the $t - V$-model to an apparent departure from the typical ground state scaling $\alpha \gtrsim 1/2$. This is discussed in App.~\ref{sec:additional_ground_state_data}. \label{fig:additional_gs_fillings}}
  \end{center}
\end{figure}

As discussed in section \ref{sec:complexity_scaling}, the complexity of the ground states is generally found to be lower than that of excited states, with the scaling coefficient $\alpha$ taking values from $\alpha \approx 1/2$ up to $\alpha \approx 0.8$. We plot data for additional filling factors in Fig. \ref{fig:additional_gs_fillings} with largely similar results, except for the half-filled $t - V$-model, which has a doubly-degenerate ground state and a significantly lower complexity. In general, a low value of $\alpha$ means that the state contains structure that is not apparent in the natural occupations due to the degeneracy. Ground states are expected to have a lower ratio $\alpha$ than excited states, because they have more two-particle correlations that restrict the available Slater determinants. The ground state of the half-filled $t - V$ - model in the limit $V \rightarrow \infty$ is an extreme example of this, as it takes (in position basis) the form
\begin{equation}
    \ket{\psi}_{\pm}=\left( \ket{010101...} \pm \ket{101010...} \right)/\sqrt{2},
\end{equation}
whose natural orbitals are the position orbitals and natural occupations are all $1/2$. This state thus actually belongs to the low-complexity class with $S_P=S_c=\log(2)$, and the scaling ansatz $S_P=\alpha N_p S_c$ does not apply. For large but finite $V$ the ground state is doubly degenerate with the eigenstates in the $\pm 1$-parity blocks approaching $\ket{\psi}_{\pm}$ as $V$ grows, and $\alpha$ approaching $0$. However, one can still form linear combinations of the degenerate ground states such that one of the components $\ket{010101...}$ or $\ket{101010...}$ is eliminated. We find that these states again closely follow the scaling Ansatz with $\alpha=1/2$ in their respective natural orbital bases.
Similar conclusion holds if the degeneracy is lifted by an additional term which discriminates between the two different charge-density wave states. Thus, the qualitative departure from the ground state scaling with $\alpha\gtrsim 1/2$ can be directly traced to the degeneracy of the $t-V$ model at half filling.

\section{Ground state complexity in a stochastic model of perturbation theory} \label{sec:hamming_dist}

\begin{figure}
  \begin{center}
    \includegraphics[width=\columnwidth]{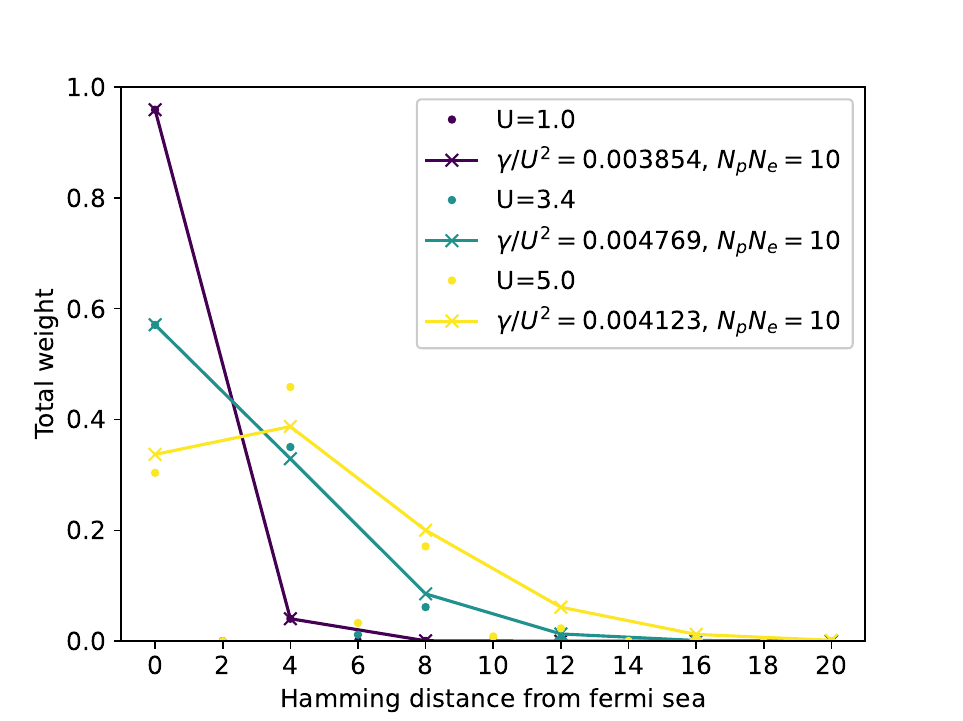}
    \caption{ Fitting
      Eqn.~\ref{eqn:hamming_from_reference} to ground state data from
      the Hubbard model. The numerical results were computed at
      half-filling for system size $L=10$ in momentum orbitals. The
      dots show the total weight of Slater configurations with a given
      Hamming distance from the Fermi sea, while the lines represent
      the model distribution. Note that the weight at Hamming
      distances not divisible by four is zero in the model of
      Eqn.~\ref{eqn:hamming_from_reference}, because all excitations
      from the Fermi sea are assumed to be pair excitations. This
      agrees with the numerical data quite well.
      \label{fig:hamming_dist}}
  \end{center}
\end{figure}

In this section we provide a heuristic model capturing essential properties of the ground states of locally interacting lattice models, and show that it leads to the scaling form $S_P=\frac{1}{2}N_p S_c$. The model is based on a picture where we have a single highest weight Slater configuration, and the weight of the other configurations decreases exponentially with the growing ``distance'' from this ``Fermi sea''.
If we fix a basis, we can think of the Slater configurations as bit strings of the occupation numbers, and measure such distances using the Hamming distance. Below we will use heuristic arguments to explicitly express the Slater weights, thus allowing us to compute $S_P$, but first we need a result that allows us to connect the correlation entropy $S_c$ to this picture.

Suppose that we draw two Slater configurations from the probability distribution $p_i$ describing a state of interest $\ket{\psi}$. The expected Hamming distance between these configurations is then
\begin{equation*}
  \left\langle x \right\rangle = \sum_{i,j=1}^{N_s} p_i p_j x_{ij},
\end{equation*}
where $x_{ij}$ is the Hamming distance between configurations $i$ and $j$. Based on the discussion in \citet{PhysRevB.103.214206}, one can express the quantity $s_c=\exp(-S_c)$ in the natural orbital basis as
\begin{equation}
  s_c=1-\frac{\left\langle x \right\rangle}{2 N_p}.
\end{equation}
For example, if we draw two configurations from a generic (Haar distributed) state, the occupied orbital positions are essentially random, and thus on average $\nu N_p$ particles of the second configuration take positions that are occupied in the first configuration. The expected Hamming distance is thus $\left\langle x \right\rangle=2 (1-\nu) N_p$, and we recover the generic state result $s_c=\nu$. This is the minimal value of $s_c$ at a fixed filling fraction, while the maximal value $s_c=1$ is obtained if $\ket{\psi}$ is a Slater state and thus $\left\langle x \right\rangle=0$.


Consider now the structure of a typical ground state in the weakly interacting limit. For zero interactions the state is a Slater determinant in the momentum basis where all orbitals below the Fermi level are occupied and all orbitals above the Fermi level are empty. Crucially, the interaction acts perturbatively by lifting \emph{pairs} of particles from below the Fermi momentum to above the Fermi momentum. For example, the interaction terms in the Hubbard model are of the form $\frac{U}{L} c^\dagger_{\uparrow k_1 - q} c^\dagger_{\downarrow k_2 + q} c_{\uparrow k_1 } c_{\downarrow k_2}$, where $L$ is the system size, and particles at momenta $k_1$ and $k_2$ are lifted to momenta $k_1+q$ and $k_2-q$. However, we cannot just use first order perturbation theory to model the limit $L \rightarrow \infty$, as that would imply that configurations with more than one pair excitation have zero weight, which precludes the linear scaling of $S_P$ with system size. Instead, we model a large system by assuming that there are $N_e N_p$ independent pair excitations, each of which occurs with a small probability $\gamma/N_e$. The $1/N_e$ scaling is required, as otherwise the number of excitations would grow with $N_e$, which may increase with system size, as the number of possible pair excitations increases faster than $N_p$. We will also assume that $\gamma$ is small, which means the excitations are rare. Thus we can assume that the excited pairs do not ``overlap'', always affecting a different set of four orbitals.

Accoring to the above assumptions the number of excited pairs is distributed binomially, with $N_p N_e$ the number of trials and $\gamma/N_e$ the success probability. For the sake of comparing to numerical data, we note that the Hamming distance $x_{fs}$ measured from the Fermi sea follows the distribution
\begin{equation}
  \begin{split}
    & P(x_{fs}=4 k) \\
    & =\left(\frac{\gamma}{N_e}\right)^k \left(1-\frac{\gamma}{N_e}\right)^{N_p N_e-k}{N_p N_e \choose k},
  \end{split}
  \label{eqn:hamming_from_reference}
\end{equation}
and $P(x_{fs})=0$ when $x_{fs}$ is not divisible by four. Fig. \ref{fig:hamming_dist} shows that a reasonable fit to Hubbard model data is obtained with $N_e=1$ and $\gamma/U^2 \sim 0.004 ... 0.005$. It is immediately clear that the assumption of non-overlapping pair excitations is correct to good accuracy, as Hamming distances not divisible by four have a very low weight. We also expect that $\gamma \sim U^2$, because amplitudes in first order perturbation theory scale proportionally to $U$ while probabilities scale as $U^2$. It would be possible to build a more refined model by taking into account that some excitations occur with higher probability than others, but we will leave this to future work.


When we draw a random state from the distribution, it has on average $N_p N_e \gamma/N_e=N_p \gamma$ excited pairs. Drawing two such states, the expected Hamming distance between them is $\left\langle x \right\rangle = 2 \cdot 4 N_p \gamma$, where the factor $2$ is because both states have $N_p \gamma$ excitations and the factor $4$ because a pair excitation causes four opposite bits. The factor $\gamma$ is thus related to $s_c$ as
\begin{equation*}
  s_c \approx 1-4\gamma.
\end{equation*}
Note that this is only correct for small $\gamma$, as otherwise the excitations start to overlap and the calculation becomes more complicated. Indeed, the lower limit for $s_c$ is $\nu$, so we should have $4\gamma \ll 1-\nu$. We can then compute $S_c$ as
\begin{equation}
  S_c = -\log(s_c) \approx 4\gamma.
\end{equation}

On the other hand, $S_P$ is related to the collision probability, which in our model means the probability that the two configurations we draw from the distribution are exactly the same, i.e. the probability that exactly the same excitations occur twice. This probability can be written as
\begin{equation}
  \begin{split}
    & P(x=0) \\
    &=\sum_{k=0}^{N_p N_e} \left( \frac{\gamma}{N_e} \right)^{2k} \left( 1 - \frac{\gamma}{N_e} \right)^{2(N_p N_e - k)} {N_p N_e \choose k} \\
    &= \left( \left(\frac{\gamma}{N_e}\right)^2 + \left(1-\frac{\gamma}{N_e}\right)^2 \right)^{N_e N_p}
  \end{split}
  \label{eqn:P_x_0_derivation}
\end{equation}
Keeping in mind the restriction to rare excitations, we expand the complexity $S_P=-\log(P(x=0))$ to linear order in $\gamma$ as
\begin{equation}
  S_P = -\log(P(x=0)) \approx 2 \gamma N_p.
\end{equation}
We thus arrive at the relation $S_P = \frac{1}{2} N_p S_c$. We note that this result can also be obtained simply by approximating $P(x=0)$ by the probability of selecting twice the unperturbed Fermi sea, corresponding to $k=0$ in Eqn.~\ref{eqn:P_x_0_derivation}, as the other contributions are higher order in $\gamma$.


The main point in this simplistic model is that $S_c \approx 1-s_c \approx 4\gamma$ is proportional to the average number of excitations per particle from the unperturbed Fermi sea, but the constant of proportionality depends on the type of the excitations. The scaling coefficient $\alpha=\frac{1}{2}$ arises because the excitations are typically pair excitations. Had we assumed single-particle excitations, we would have obtained $S_c \approx 1-s_c \approx 2\gamma$, and the end result would have been $S_P=N_p S_c$.

\section{Computational details} \label{app:detail}

We perform exact diagonalization calculations using the QuSpin package \cite{10.21468/SciPostPhys.2.1.003,10.21468/SciPostPhys.7.2.020}, which allows easy building of Hamiltonian matrices for the fermionic lattice models considered here. The package also allows selecting specific symmetry sectors of lattice models, fixing e.g. quantum numbers corresponding to center-of-mass momentum and parity under reflection $p \rightarrow -p$. For the excited state calculations we perform full diagonalization of the selected symmetry block using standard dense hermitian methods, while for the ground state results we employ ARPACK-based sparse methods included in the QuSpin library and Scipy \cite{2020SciPy-NMeth}.

To study the entropy $S_{P_\mathcal{B}}$ in different bases $\mathcal{B}$, we need to change the single-orbital basis for the full many-body eigenstates which are initially computed in the position basis. In the second quantized formalism, an orbital transformation for a system with $N_o$ orbitals is specified by an $N_o \times N_o$ unitary matrix $U$ acting on the annihilation operators as
\begin{equation}
  \vec{c'}=U\vec{c}, ~ \vec{c}=[c_1,c_2 \cdots, c_{N_o}]^T.
\end{equation}
The unitary matrix can be parametrized by a hermitian matrix $A$ such that $U=\exp(i A)$, and this transformation can then be expressed as an operator $\hat{U}$ in the many-body Fock space as
\begin{equation}
  \hat{U}=\exp(i \vec{c}^\dagger A \vec{c}),
\end{equation}
acting on operators as $\hat{U}^\dagger \vec{c} \hat{U}= U \vec{c}$. That the orbital rotations can be expressed in such exponentiated form is referred to as the Thouless theorem \cite{thouless1960} in the literature \cite{PhysRevLett.120.110501}.

The operator $\hat{A}=\vec{c}^\dagger A \vec{c}$ is represented as a sparse matrix that only couples basis states connected by a single hop, thus having $N_p N_h Q$ non-zero elements, where $N_p$ and $N_h$ are the number of particles and holes, respectively, and $Q$ is the number of Fock basis states. Applying the operator $\hat{U}$ on a state in the Fock space can be carried out by sparse matrix methods, where the only large matrix operation is matrix-vector multiplication by $\hat{A}$. \cite{al2011computing_action_of_matrix_exponential,higham2010_computing_matrix_functions} For small systems, the non-zero matrix elements of $\hat{A}$ can be computed and stored in memory in a sparse matrix format. For large systems, it is advantageous to compute the matrix elements of $\hat{A}$ on the fly when performing the matrix-vector multiplication, as memory access becomes the bottleneck of the computation.

For basis optimization we again parametrize the orbital basis in the form $U=\exp(iA)$ and perform a conjugate gradient minimization of the Renyi entropy $S_{P_\mathcal{B}}$ with the elements of the hermitian matrix $A$ treated as free parameters. For the single-component models, we use a random matrix $U \sim \text{CUE}(N_o)$ as the starting point of the minimization, with $N_o$ the number of orbitals in the model. For the two-component Hubbard model, we enforce component conservation meaning that $A$ is block-diagonal and does not mix different spin components.

\bibliography{complex.bib}

\end{document}